# Title: Dispersive temporal holography for single-shot recovering comprehensive ultrafast dynamics

**Authors:** Wenchao Wang, Tianhao Xian, and Li Zhan *

**Affiliations:**
State Key Laboratory of Advanced Optical Communication Systems and Networks, School of Physics and Astronomy, Shanghai Jiao Tong University, Shanghai, China.

*Corresponding author. Email: lizhan@sjtu.edu.cn

**Abstract:** It is critical to characterize the carrier and instantaneous frequency distribution variation in ultrafast processes, all of which are determined by the optical phase. Nevertheless, there is no method that can single-shot record the intro-pulse phase evolution of pico/femtosecond signals, to date. By analogying holographic principle in space to the time domain and using the time-stretch method, we propose the dispersive temporal holography to single-shot recover the phase and amplitude of ultrafast signals. It is a general and comprehensive technology and can be applied to analyze ultrafast signals with highly complex dynamics. The method provides a new powerful tool for exploring ultrafast science, which may benefit many fields, including laser dynamics, ultrafast diagnostics, nonlinear optics, and so on.

**Main Text:**

### Introduction

One of the most fundamental and challenging problems throughout ultrafast science is to fully characterize the complex electric fields of the ultrashort optical pulses. Nowadays, in laboratories and industry, the techniques called FROG (Frequency-Resolved Optical Gating) ([1, 2]), SPIDER (Spectral Phase Interferometry for Direct Electric-field Reconstruction) ([3]), and IAC (Interferometric Autocorrelation) ([4]) are successfully used for pulse reconstruction. However, in some crucial explorations (such as how the ultrafast and nonlinear processes occur and evolve), the recordings must be genuine real-time and accurate. The above mainstream correlation-based methods require multi-point scanning to determine the phase and cannot track the ultrafast signals consecutively. Even if the single-shot detection is realized ([5]), the nonlinear-effect-dependent measurements must satisfy some certain conditions, such as phase matching or pretty high pulse peak power. That means such reconstructions, or to say a kind of estimation, are affected by the properties of the signals themselves. It may have considerable errors to reveal the irregular pulse waveforms has a single occurrence.

Recently, a technique independent of nonlinear effects called dispersive Fourier transform (DFT) ([6, 7]) has made it possible to real-time detect spectral evolution in complex ultrafast systems ([8, 9]). Utilizing their spectral interferometry, the relative carrier phase and the time interval limited at optical soliton molecules can also be retrieved ([10, 11]). Fast dynamics such as the mode-locking buildup ([8]), breathers ([9]), supercontinuum ([12]), pulsating solitons ([13]) were intensely studied using DFT-based single-shot spectral measurement in recent years. However, the method cannot describe the panorama enough. It missed the complex waveform shapes and the temporal phase distribution, and failed to directly reveal the intrinsic physical mechanism by the observations.

The basic behavior of many vital nonlinear dynamics is still unknown because of lacking real-time time-domain measurements. A typical example is the optical rogue wave (*14*), which like ocean rogue wave (*15*), always appears and disappears suddenly. In theory, it was closely related to the breathers (*16, 17*). They have a strong connection with the Fermi-pasta-Ulam-Tsingou (FPUT) paradox (*18*), and the modulation instability (*19*). Previous researches only report the time-stretch spectroscopy or the oscilloscope traces, but the transient phase evolution is more significant for determining the origin of rogue waves. Therefore, it is urgent and essential to developing a single-shot phase and amplitude reconstruction technology for detect complex ultrafast dynamics.

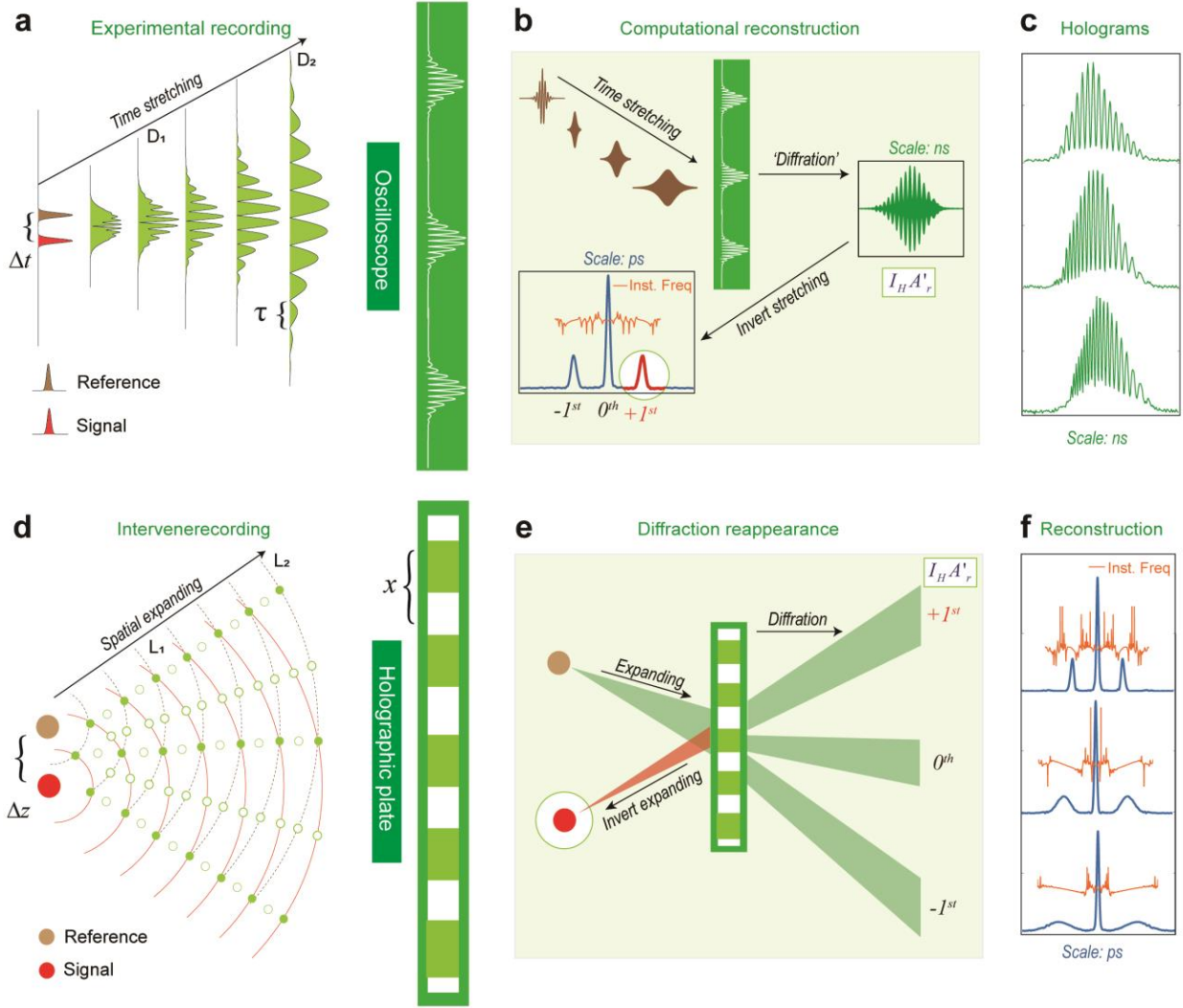

**Fig. 1. The basic idea of dispersive temporal holography.** (**a**) The signal and reference pulses are stretched by a large amount of dispersion, and the formed interference fringes are recorded by an oscilloscope. (**b**)The information in the fringes can be retrieved by the reference beam. Corresponding to (**d**) "intervene recording" and (**e**) "diffraction reappearance" in classical holography. (**c**) Holograms of signal pulses with different chirps. (**F**) The corresponding reconstructions.

**Principle of DTH**

The holography *(20)* is a technology that can reconstruct the 3-D image from 2-D holograms, which is realized by recovering both the intensity and the phase in physics. Another feature is that, it transfers the measurement of the signal to the holograms, making the needed sampling rate no longer depends on the signal itself *(21)*. Applying such a principle on ultrafast signals can restore the intensity or phase in the hologram and vastly reduce the sampling requirements. The proposed method (Figs. 1a and b) has a one-to-one map with the classical spatial holography (Figs. 1d and e). In spatial holography, the signal and the reference can be regarded as point sources, and the interference patterns are distributed along the divergent propagation. The interference period increases with the propagation distance $L$ and inversely proportional to the initial spacing $\Delta z$. Similarly, the signal and the reference pulses in DTH (assuming their spectra are overlapped) can be considered "point sources" in the time domain. Akin to the spatial propagation, applying the time-stretch (TS) technology to the ultrafast signals can create temporal interference fringes. The period is determined by two factors: the dispersion $D$ and the time interval $\Delta t$. Different from restoring the spatial holograms by holographic plates, the temporal "holograms" are recorded by oscilloscopes.

In spatial holography, the so-called Nyquist sampling frequency $F_{Ns}$ *(21, 22)* is determined by the interference period, and it can be controlled by spatial expanding. As shown in Fig. 1d, the holographic plate for measurements at a more distant location $L_2$ should be larger in size, but it can be lower-resolution, comparing with $L_1$. The same principle applies to the DTH, and TS plays the same role as the spatial expanding. If we define time-stretching ratio as $g = D_2 / D_1$, $F_{Ns}$ can be reduced by $g$ times. When TS is enough, the oscilloscope with a several-GHz sampling rate can record the holograms of ultrafast signals without distortion.

In holographic principle, the outgoing light $A_{out}(z,T)$ is expressed as:

$$A_{out}(z,T) = I_H A'_r = \left(|A_r|^2 + |A_s|^2\right)A'_r + A_s A^*_r A'_r + A^*_s A_r A'_r, \tag{1}$$

where $I_H$ represents the intensity distribution of the hologram. $A_s(z,T)$, $A_r(z,T)$ and $A'_r(z,T)$ are the complex amplitude of the signal, recording and reappearance references. $z$ and $T$ are the space and time coordinates. Three terms on the right-hand side of the equation correspond to the $0^{th}$, $+1^{st}$, and $-1^{st}$ order diffraction light. If $A'_r(z,T)$ is exactly the same as $A_r(z,T)$, the second term is expressed as $|A_r|^2|A_s|e^{i\varphi_s}$, and thus the signal can be exactly recovered.

Since the above equations should not certainly depend on time or space coordinates, the holographic principle is actually applicable to the time dimension. The "diffraction reappearance" step of DTH shown in Fig. 1b is computer generated. The reference signal $A'_r(z,T)$ after TS can be expressed as $A'_r(z,\omega)e^{iD\omega^2/2}$, where the $A'_r(z,\omega)$ is the frequency-domain description, $D$ is the total dispersion used for stretching. On the other hand, the sampling rate of $I_H$ should ascends to the same level with $A'_r(z,T)$, it is achieved by zero-padding in the frequency domain *(22)*. Then, the outgoing light after diffraction is $I_H A'_r$. Just like the reappeared virtual object image after diffraction is spatially reversed, we can obtain the recovery signal by removing the introduced dispersion to realize the temporal inversion. The $+1^{st}$ order signal is $|A_r|^2|A_s|e^{i\varphi_s}$, that means we get the all-round information, including the pulse shape, phase, and duration. If $A'_r(z,T)$ is a constant $C$, the $+1^{st}$ order term is expressed as $CA^*_r A_s = C|A_r A_s|e^{i(\varphi_s - \varphi_r)}$, and the relative phase can be retrieved.

In spectral interferometry, or FROG and SPIDER, the expansion terms are consisting of convolution or correlation functions in time-domain, requiring multi-point scanning to reconstruct the temporal phase. However, in DTH, the expansions are the products of the temporal terms, which can directly describe the signal. Hence, the single-shot measurement can be achieved. Besides, in spectral interferometry, the relative carrier phase and time interval are retrieved by the interference fringes location and spacing (*10, 11*). In principle, any phase change does exert influence on the fringes. For instance, if the signal is chirped, the fringes will be sparse on the one side and dense on the other side. As shown in Figs. 1c and f, holograms in more chirped cases tend to be more unevenly distributed. Like holography can decode the spatial hologram patterns, DTH provides a method to extract the hidden information.

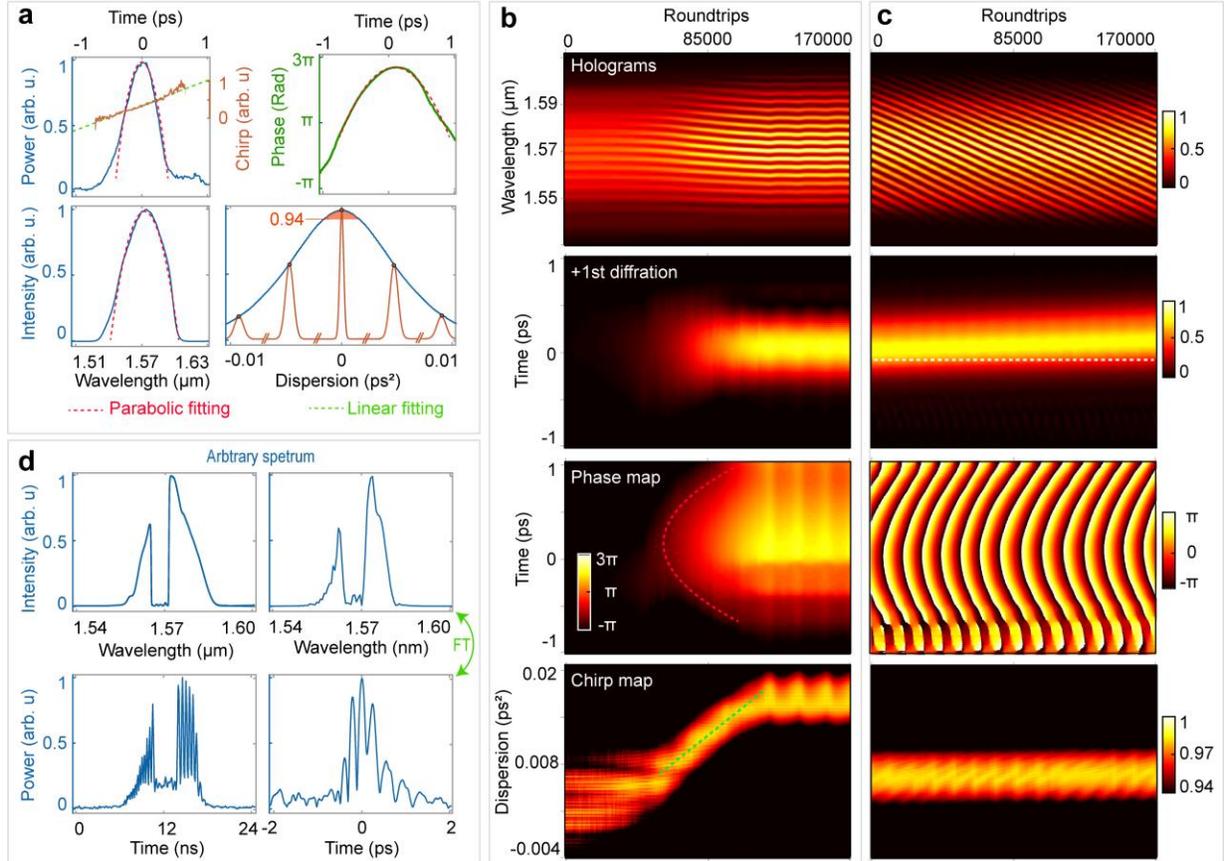

**Fig. 2. Replication experiments.** (**a**) Top left: The reconstructed amplified pulse with linear chirp. Top right: The corresponding phase. Bottom left: The measured spectrum of the seed parabolic pulse. Bottom right: Explanation of the chirp map. (**b**) Top: The holograms of the SSA, measured by an oscilloscope. The rests: The retrieved pulse waveforms, phase, and chirp of successively 170000 roundtrips, according to the holograms. (**c**) The results of moving the optical delay line in the signal arm. (**d**) To test the proposed method in case of arbitrary spectrum. Top left: The spectrum measure by an optical spectrum analyzer. Bottom left: The holograms. Top right: The Fourier transformation of the retrieved pulse. Bottom right: The reconstructed pulse.

**Replication experiments**

To test the proposed method, we use an ultrafast Er-doped fiber laser (*26*) as the source, and divide it to two output paths by an optical coupler with a 50:50 split ratio. One path experiences certain events, while the other serves as the reference.

The first event (Fig. s1) is the self-similar amplification (SSA) of parabolic pulses *(23)*. We reconstructed the pulse after amplification, the pulse width is ~450 fs, as shown in the top of Fig. 2a. It has a linear chirp and a parabolic phase distribution, in agreement with theoretical prediction *(24)* and the FROG traces *(25)*. The optical spectrum of seed pulse is shown in Fig. 2a bottom left, and it can also be parabolic fitted. The holograms in Fig. 2b show the power and phase change. As the amplification, the reconstructed pulse shape is near parabolic evolved, as is shown below. Because the complex amplitude of the light field is described as $A(T) = |A(T)|e^{i\varphi(T)+i\omega_0 T}$, it has a carrier modulation and causes oscillation in the phase. In order to visualize the phase evolution, what we draw in Fig. 2b is $\varphi(T)$. The phase always keeps a parabolic shape in the SSA process. No matter at normal or abnormal dispersion, the chirp-free pulse will be broadened, and the peak power will be decreased, as shown at the bottom of Fig. 2a. We can visualize the chirp variation as long as recording the peak power change. In this way, we draw the chirp map at the bottom right of Fig. 2b. The linear chirp is gradually accumulated during SSA, because the Kerr effect in SSA brings the additional intensity-dependent phase shift and positive chirp *(26)*. For comparison, we carried out the numerical simulations (Fig. s2). The nonlinear phase shift of nearly 4π is generated in the center of the pulse with the chirp of $\sim 0.0016\, ps^2$, close to the experimental values.

The second event (Fig. s3) is the moving of an optical delay line in the signal arm (Fig. 2c). The interference fringes of the hologram indicate the relative phase has a shift, and the restored signals' temporal position has a delay of $\sim 75\, fs$ in total, in agreement with the carrier phase shift (~30 π) in the phase diagram. It shows that the method can accurately reveal the phase and group velocity respectively. The retrieved chirp in the bottom diagram stays no change because the signal properties keep unchanged.

In the last event (Fig. s4), we filter out the middle $16\, nm$ of the signal to test the method applying to arbitrary spectrum pulses. According to the holograms (Fig. 2d bottom left), we get the retrieved pulse shape, as shown in the bottom right of Fig. 2d. We find its Fourier transform spectrum (Fig. 2d top right) is the same as the original spectrum in the top left of Fig. 2d. It demonstrates the feasibility of measuring complex ultrafast signals, which is the inherent advantage of holography *(20)*, independent of the specific characteristic of the signals. Besides, we should note that the proposed method is a direct measurement instead of an estimation because it does not depend on any intermediate nonlinear processes *(5)*.

**Conclusion**

We propose the DTH technique to reconstruct the phase and amplitude in ultrafast dynamics, which is a general and comprehensive single-shot ultrafast measurement method, and suitable for analyzing both phase and amplitude of ultrafast signals. Much rather than a supplementary for FROG, or SPIDER, it can also extract the phase and amplitude inside the pulse, and can be used in ultrafast dynamics, chemical reactions *(39)*, attosecond science *(40)*, etc.

**References:**

1. R. Trebino, F. R. O. Gating, *Frequency-Resolved Optical Gating: the Measurement of Ultrashort Laser Pulses* (Kluwer Academic, 2002).

2. R. Trebino, K. DeLong, D. Fittinghoff, J. Sweetser, M. Krumbügel, B. Richman. Measuring ultrashort laser pulses in the time-frequency domain using frequency-resolved optical gating. *Review of Scientific Instruments.* **68**, 3277-3295 (1997).

3. C. Iaconis, I. A. Walmsley, Spectral phase interferometry for direct electric-field reconstruction of ultrashort optical pulses. *Optics. Letters.* **23**, 792-794 (1998).

**Acknowledgments:**

National Natural Science Foundation of China grant 11874040 (PV, CHO)

Foundation for leading talents of Minhang, Shanghai.

**Author contributions:**

L. Z. proposed the conception. L. Z., and W.W. designed and interpreted the results of all experiments. W. W. performed all of the experiments and simulation, analyzed the results. T. X. provided advices. W.W. wrote the paper. L. Z. review and editing the paper.

**Competing interests:** Authors declare that they have no competing interests.

**Materials & Correspondence:** Correspondence and requests for materials should be addressed to L. Z.

**Data availability**

All data in the main text or the supplementary materials are available from the corresponding authors on reasonable request. Source data are provided with this paper.

**Supplementary Materials**
    Figs. S1. Experimental setup of event 1.
    Figs. S2. Numerical simulation of event 1.
    Figs. S3. Experimental setup of event 2.
    Figs. S4. Experimental setup of event 3.

Figs. S5. Periodic spectral evolutions at the noise-like regime, recorded by DFT.

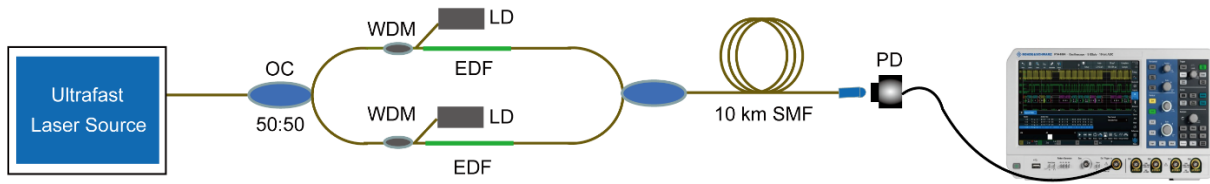

**Fig. S1. Experimental setup of event 1.**

OC: optical coupler. WDM: wavelength division multiplexer. EDF: Er-doped fiber. LD: laser Diode, centered at 980 nm. SMF: standard single mode fiber. PD: Photodetector. The ultrafast laser source is a Er-doped fiber laser mode-locked via nonlinear polarization rotation technology and delivers ~40nm parabolic pulse centered at 1.57 mm. The 10km SMF is the dispersive medium to achieve time stretching. Finally, the outputs are recorded by an oscilloscope.

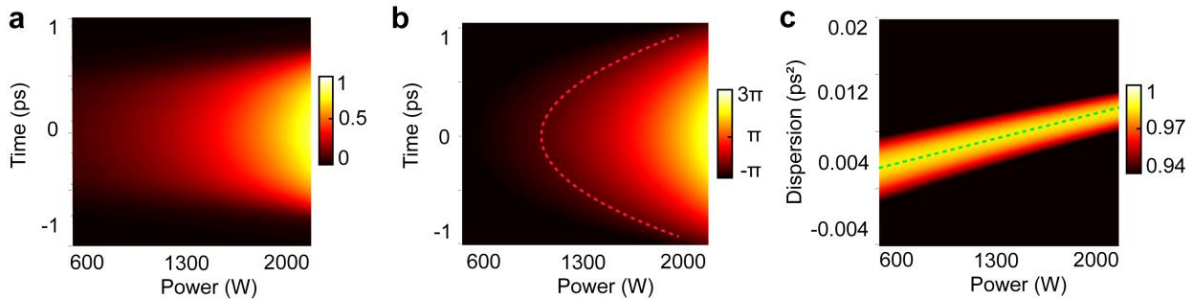

**Fig. S2. Numerical simulation of event 1.**

(**a**) The pulse waveforms. (**b**) The phase map. (**c**) The chirp map. The simulation parameters are set to be consistent with the experiment.

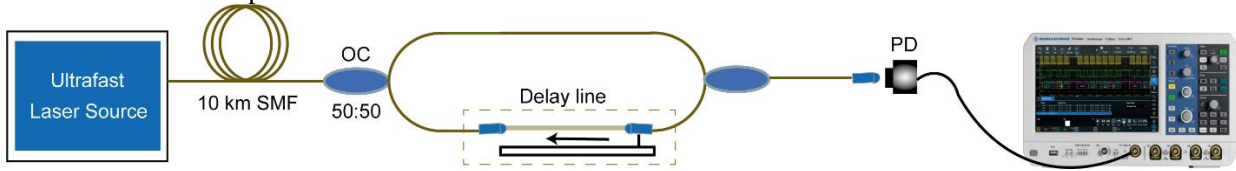

**Fig. S3. Experimental setup of event 2.**

The delay line is a commercial optical delay line and it is motor controlled.

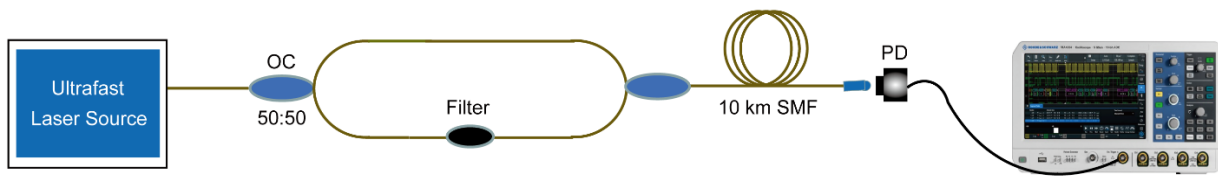